\documentclass{article}
\usepackage{graphicx,latexsym,amsmath,amsfonts,amssymb,epsfig,theorem}

\newcommand{\ZZ}{\mathbb{Z}}
\newcommand{\co}{\cal{O}}
\newcommand{\ket}[1]{{|#1\rangle}} 

\newtheorem{lemma}{Lemma}

\author{John Proos \quad and \quad Christof Zalka \\ \\
Department of Combinatorics and Optimization \\
University of Waterloo, Waterloo, Ontario \\
Canada N2L 3G1 \\
\small e-mail: {\tt japroos@math.uwaterloo.ca \quad zalka@iqc.ca }}

\title{Shor's discrete logarithm quantum algorithm for elliptic curves}

\begin{document}        

\maketitle

\begin{abstract}
We show in some detail how to implement Shor's efficient quantum
algorithm for discrete logarithms for the particular case of elliptic
curve groups. It turns out that for this problem a smaller quantum
computer can solve problems further beyond current computing than for
integer factorisation. A 160 bit elliptic curve cryptographic key
could be broken on a quantum computer using around 1000 qubits while
factoring the security-wise equivalent 1024 bit RSA modulus would
require about 2000 qubits. In this paper we only consider elliptic
curves over GF($p$) and not yet the equally important ones over
GF($2^n$) or other finite fields. The main technical difficulty is to
implement Euclid's gcd algorithm to compute multiplicative inverses
modulo $p$. As the runtime of Euclid's algorithm depends on the input,
one difficulty encountered is the ``quantum halting problem''.

\end{abstract}

\tableofcontents

\section{Introduction} \label{intro}
In 1994 Peter Shor presented two efficient quantum algorithms
\cite{shor} for computational problems for which no polynomial time
classical algorithms are known. One problem is to decompose a (large)
integer into its prime factors. The other problem, which we consider
here, is finding discrete logarithms over finite groups.  The
classical complexity of this problem seems to depend strongly on the
underlying group. A case for which (known) classical algorithms are
particularly inefficient are elliptic curve groups defined over finite
fields. Actually most public key cryptography in use today relies
either on the presumed hardness of integer factoring (RSA) or that of
discrete logarithms over finite fields or elliptic curves.

Elliptic curve cryptography (ECC) is sometimes preferred because it
allows shorter key sizes than RSA. This is because the best classical
integer factoring algorithms (the number field sieve, see
e.g. \cite{nfs}), although superpolynomial, have less than exponential
complexities. Very roughly the complexity is $O(e^{c \log^{1/3}n})$,
where $n$ is the integer to be factored. On the other hand, for
discrete logarithms over elliptic curves, nothing better than
``generic'' algorithms are known, thus algorithms which work for any
group. These algorithms, e.g. the Pollard $\rho$ algorithm \cite{rho},
have truly exponential complexity.

Shor's quantum algorithms for integer factoring and discrete
logarithms have about equal complexity, namely typically
$O(n^3)$. Thus there is a larger complexity gap between classical and
quantum for discrete logarithms than for factoring.

Proposals have been made \cite{beau,zalka} for optimised
implementations of the quantum factoring algorithm, in particular for
minimising the number of qubits needed. The best current result by
S.Beauregard \cite{beau} is that about $2 n$ qubits are enough. We
attempt here a similar optimisation for discrete logarithms over
elliptic curves. The implementation is more difficult, but we still
get an algorithm that uses less qubits and time to solve a problem of
similar classical difficulty when compared to factoring.  For problems
that can now barely be solved, the number of qubits is not much less
than for factoring, but in the future, with more powerful classical
computers, the gap will increase.

Elliptic curves used in cryptography \cite{ansi,ieee,NIST} are defined
either over the field of arithmetic modulo a prime, thus $GF(p)$, or
over $GF(2^n)$. For our implementation we need to do arithmetic
operations in these fields, in particular we must compute
multiplicative inverses. For $GF(p)$, this is done with Euclid's
algorithm for computing the greatest common divisor (gcd), or rather
the extended version of this algorithm. This algorithm can be adapted
to the case of any finite field $GF(p^n)$, but for $n>1$ there is the
added concern of deciding how the elements of the field will be
represented.  So in this paper we only consider elliptic curves over
$GF(p)$.

Still, the implementation of the extended Euclidean algorithm is the
main technical difficulty we encounter. Fortunately, the algorithm can
be made piecewise reversible, so that not too much ``garbage'' has to
be accumulated. As for the factoring algorithm, it is possible to run
the whole algorithm with $O(n)$ qubits. For our implementation of
Euclid's algorithm to achieve the classical time complexity of
$O(n^2)$, it is necessary to terminate the steps in the algorithm at
different points, depending on the input. This is difficult to achieve
with acyclic circuits (which are necessary for computations in
``quantum parallelism''). We will relegate some of the more cumbersome
technical aspects of our solution to an appendix, and will also
discuss possible other approaches.

In trying to optimise our implementation, we were guided by practical
considerations, although to do this, one would really have to know how
an actual quantum computer will look. We put most emphasis on
minimising the number of qubits, but also on the total number of
gates. We assume that whatever can be computed classically, should be
done classically as long as it doesn't take an unreasonable amount of
computation. Basically we are trying to optimise a quantum circuit
where a gate can act on any pair of qubits, but it turns out that most
gates are between neighbours like in a cellular automaton. In contrast
to the earlier papers \cite{beau,zalka} optimising the quantum
factoring algorithm, we have not thought about parallelising the
algorithm, although this may well be of interest for actual
implementations.

\section{Review of the quantum algorithm for discrete logarithms}

\subsection{The discrete logarithm problem (DLP)} \label{DLP}

Let $G$ be a finite cyclic group and let $\alpha$ be a generator of
$G$. The discrete logarithm problem over $G$ to the base $\alpha$ is
defined as given an element $\beta \in G$ determine the unique $d \in
[0,|G|-1]$ such that $\alpha^d = \beta$. The integer $d$ is denoted by
$\log_\alpha \beta$. Note that while $G$ may be a subgroup of a
non-abelian group, $G$ being cyclic is always an abelian
group. Usually it is assumed that the order of $G$ is known.

There are two general types of algorithms for solving DLPs. The first
type, called generic algorithms, work for any group, as long as we
have a (unique) representation of group elements and we know how to
carry out the group operation. The best known classical generic
algorithms have complexity equal to about the square root of the order
of the group. Thus they are exponential in the number of bits
necessary to describe the problem.

The second type of algorithms are the algorithms which rely on
specific properties of the group or its representation. As shown in
the examples below, some groups have group specific algorithms which
can solve the DLP in subexponential or even polynomial time.

\subsubsection{Examples ($\ZZ_N$ and $\ZZ^*_p$)}

Let $N$ be a positive integer and consider the case when $G= \ZZ_N$
the additive group of integers modulo $N$. Here the generators of the
group are precisely the $\alpha \in G$ such that $\gcd(\alpha,N) = 1$
and the equation $d \cdot \alpha \equiv \beta \pmod{N}$ can be solved
by finding the multiplicative inverse of $\alpha$ modulo $N$ with the
extended Euclidean algorithm. Thus for this group the DLP can be
solved in polynomial time ($O({\log_{2}}^2{N})$).

There are however groups for which the DLP is not so easy. Suppose
that $G= \ZZ^*_p$ the multiplicative group modulo $p$, which is
cyclic, and that $\alpha$ is a generator of $G$. Then the DLP is
equivalent to solving the equation $\alpha^d \equiv \beta
\pmod{p}$. There are no known classical algorithms which can solve
this problem in polynomial time. Still, like for integer factoring,
the best algorithms have a subexponential complexity.

Note that if $G$ is a finite cyclic group of order $N$ then $G$ is
isomorphic to $\ZZ_N$ in which the DLP is easy. Thus it is not the
structure of a group, but its representation, which can make its DLP
difficult.
  
\subsubsection{Discrete logarithms over elliptic curves}

Elliptic curves over $GF(p^n)$ are finite abelian groups. Given a
point $\alpha$ on an elliptic curve we can consider the difficulty of
solving the DLP in the cyclic subgroup generated by $\alpha$. For
general elliptic curves (trace not equal to zero or one) the DLP seems
to be computationally quite hard. In particular for these curves it is
not known how to exploit the representation of the group to help solve
the DLP.  Thus the best known classical algorithms for the DLP on
these elliptic curves are the generic algorithms whose running times
are exponential in the number of bits necessary to describe the
problem.  This presumed classical hardness makes the groups useful for
cryptography and has led to systems based on these group being
included in ANSI, IEEE and FIPS standards \cite{ansi,ieee,NIST}.

\subsection{Shor's quantum algorithms}

Both of Shor's algorithms have later been understood as special cases
of a more general framework, namely the abelian hidden subgroup
problem (see e.g. \cite{ettinger,jozsa,kitaev}). While in the
factoring algorithm we are looking at subgroups of the group of
integers, $\ZZ$, in the discrete logarithm case, subgroups of $\ZZ^2$
play a role. In particular we are looking at sublattices of the
lattice $\ZZ^2$, thus elements which can be written as integer linear
combinations of two (linearly independent) vectors in $\ZZ^2$. Thus in
a way the discrete logarithm algorithm can be viewed as a 2
dimensional version of the factoring algorithm.

\subsubsection{The order finding algorithm (factoring)} \label{order}

The basis of the integer factoring algorithm is really an order
finding algorithm, which we briefly review here. We are given an
element $\alpha$ in a (finite) group $G$ and want to find its order.
That is, the smallest non-negative integer $r$ with $\alpha^r = e$,
where $e$ is the neutral element. To do this, we prepare a (large)
superposition of $N$ ``computational'' basis states $\ket{x}$ and
compute $\alpha^x$ in ``quantum parallelism'':
$$ \frac{1}{\sqrt{N}} \sum_{x=0}^{N-1} \ket{x} \qquad \to \quad 
\frac{1}{\sqrt{N}} \sum_{x=0}^{N-1} \ket{x,\alpha^x} $$
Where $N$ is much larger than any order $r$ that we expect. Now
imagine that we measure the second register and get $\alpha^{x_0}$
(the measurement is not actually necessary). Then the first register
will be left in a superposition of the form
$$ c \cdot \sum_{k=0}^{\approx N/r} \ket{x_0+k \cdot r} $$
where $x_0$ is a random number between $0$ and $r-1$. Now a quantum
Fourier transform (of size $N$) will leave this register in a
superposition dominated by basis states that are close to multiples of
$N/r$. Thus a measurement will yield such a state with high
probability. If $N$ is chosen larger than the square of any expected
order $r$, it is possible to calculate $r$ from the observed state
with high probability. Also $N$ is chosen to be a power of 2, as this
gives the simplest ``quantum fast Fourier transform'' (QFFT).

\subsubsection{Assumption for discrete log: order is prime (and known)}

First let us justify a simplifying assumption that we make. We assume
that the order of the base $\alpha$ of the elliptic curve discrete
logarithm is prime and that we know this prime. This is true for the
cases standardised for cryptographic use \cite{ansi,ieee,NIST}.  Also,
if we don't know the order of $\alpha$, we can find it with the above
order finding algorithm and also decompose it into its prime factors
with the integer factoring algorithm. Then there is a standard way to
reduce the DLP in a group with composite order, $N$, into several DLPs
with orders equal to the prime factors of $N$ (see \cite{pq}).  Thus
our simplifying assumption is really without loss of generality.

\subsubsection{The discrete logarithm algorithm} \label{dlog}

So we have $\alpha^q = e$, with $q$ prime and $\beta= \alpha^d$ where
$d$ is unknown and between $0$ and $q-1$. Consider the function
$f(x,y) = \alpha^x \beta^y$ for integers $x$ and $y$. This function
has two independent ``periods'' in the plane $\ZZ^2$, namely
$$f(x+q,y) = f(x,y) \qquad \mbox{and} \qquad f(x+d,y-1) = f(x,y)$$
Thus all $x,y$ with $f(x,y) = e$ define a sublattice of $\ZZ^2$. The 2
dimensional Fourier transform then leads to the dual lattice from
which $d$ can be determined.  Note that $f(x,y)$ can be thought of as
being defined over ${\ZZ_q}^2$ as
$f(x,y)=f(x~\mbox{mod}~q,y~\mbox{mod}~q)$.

For didactic purposes let us first imagine that we knew a way to carry
out the quantum fast Fourier transform of order $q$ (QFFT$_q$) , as
then the algorithm would be particularly nice. (Actually it has been
shown how to do this approximatively \cite{kitaev,hallgren}, but we
won't use these constructions.) Then we start with the following state
of two quantum registers, and compute $\alpha^x \beta^y$ in ``quantum
parallelism'':
$$ \frac{1}{q} \sum_{x=0}^{q-1} \sum_{y=0}^{q-1} \ket{x,y} \qquad \to
\quad \frac{1}{q} \sum_{x=0}^{q-1} \sum_{y=0}^{q-1} 
\ket{x,y,\alpha^x \beta^y} $$
Again, imagine that we now measure the last register (although this is
again not necessary). Then we will obtain a random element
$\alpha^{x_0}$ of the group generated by $\alpha$, where $x_0$ is
between $0$ and $q-1$. We will then find the first two registers in a
superposition of all $x,y$ with
$$ \alpha^x \beta^y = \alpha^x (\alpha^d)^y =\alpha^{x_0} $$
Because the order of $\alpha$ is $q$, this is equivalent to
$$ x + d y \equiv x_0  \qquad (\mbox{mod} ~q) $$
Or equivalently $x = (x_0-d y)~\mbox{mod}~q$. Thus for each $y$ there 
is exactly one solution, and so the state of the first two registers is:
$$ \frac{1}{\sqrt{q}} \sum_{y=0}^{q-1} \ket{x_0 - d y,y} $$
Now we Fourier transform each of the two registers with our
(hypothetical) quantum Fourier transform of order $q$, which acts on
basis states as
$$ \ket{z} \qquad \to \quad \frac{1}{\sqrt{q}} \sum_{z'=0}^{q-1} 
\omega_q^{z z'} \ket{z'} \qquad \mbox{where} \quad 
\omega_q = e^{2 \pi i/q} $$
We obtain
$$ \frac{1}{\sqrt{q}} \frac{1}{q} \sum_{x',y'=0}^{q-1}~ 
\sum_{y=0}^{q-1} \omega_q^{(x_0-d y)x'} \omega_q^{y y'} \ket{x',y'}
$$
The sum over $y$ is easy to calculate. It gives $q\omega_q^{x_0 x'}$
if $y' \equiv d x' (\mbox{mod} ~q)$ and vanishes otherwise. Thus we
get:
$$ \frac{1}{\sqrt{q}} \sum_{x'=0}^{q-1} \omega_q^{x_0 x'} 
\ket{x',y'= d x' \mbox{mod} ~q} $$
We now see that the probability of measuring a basis state is
independent of $x_0$, thus it doesn't matter which $x_0$ we measured
above. By measuring, we obtain a pair $x',y'$ from which we can
calculate $d= y' (x')^{-1} \mbox{mod} ~q$ as long as $x' \not = 0$.
(The only disadvantage of allowing the order $q$ not to be a prime,
would be that we would require $\gcd (x',q)=1$.)

\subsubsection{Using a Fourier transform of order $2^n$ instead of $q$}

In practise we will want to replace each of the two QFFT$_q$'s with a
quantum fast Fourier transform of order $2^n$ (QFFT$_{2^n}$), because
this is easy to implement. For the QFFT$_q$ above we will always
obtain a pair $x',y'$ with $y' \equiv dx' \mbox{mod} ~q$ in the final
measurement.  However, for the QFFT$_{2^n}$ we will get a high
probability of measuring a pair $x',y'$ if
$$ (x' q/2^n,y' q/2^n) \approx (k, d k)  \qquad \mbox{for some} ~k $$
For $2^n \approx q$ we have a good (constant) probability of getting
the right values in ${\ZZ_q}^2$ by rounding. In appendix \ref{FT A} we
make this analysis in detail and show that by investing a reasonable
amount of classical post-processing we can with probability close to 1
obtain the discrete logarithm with a single run of the quantum
algorithm. (Of course this is on a perfect, noise free, quantum
computer...)  More classical post-processing increases the chances of
success because now we can try out several values in the vicinity of
the values $x',y'$ which we measured. Also there is a tradeoff between
$n$, thus the number of qubits, and the success probability. For $n$
increasing beyond $\log_{2} q$ the probability of failure decreases
exponentially.

\section{Elliptic curves}

As mentioned earlier, elliptic curves over finite fields form abelian
groups.  We will now present a brief introduction to elliptic curves
over fields of characteristic not equal to $2$ or $3$ (i.e. $1+1\neq
0$ and $1+1+1 \neq 0$). For a more in depth introduction to elliptic
curves and their use in cryptography see ~\cite{cha,kob,men}.

Let $K$ be a field of characteristic not equal to $2$ or $3$. An
elliptic curve over $K$ is the set of solutions $(x,y) \in K \times K$
to the equation
\begin{equation} \label{eqnec1}
E: y^2 = x^3+ax+b
\end{equation}
where $a,b \in K$ are constants such that $4a^3 +27b^2 \neq 0$,
together with the point at infinity, which is denoted $\co$. The
solutions to equation \ref{eqnec1} are called the finite points on the
curve and together with $\co$ are called the points on the curve. We
will use $E$ to denote the set of points on an elliptic curve.

The group operation on the points is written additively and is defined
as follows. If $P \in E$ then $P+ \co = \co +P = P$. If $P=(x_1,y_1),
R=(x_2,y_2) \in E$ then
\begin{equation} \label{ECop}  
P+R = 
\begin{cases}
 \co & \textrm{if $(x_2,y_2) = (x_1,-y_1)$,}\\
 (x_3,y_3 ) & \textrm{otherwise,}
\end{cases} 
\end{equation} 
where $x_3 =\lambda^2-(x_1+x_2)$, $ y_3=\lambda (x_1-x_3)-y_1$, 
\[
\lambda = \begin{cases}
(y_2-y_1)/(x_2-x_1) & \textrm{if $P \neq R$} \\
(3x_1^2+a)/(2y_1) & \textrm{if $P = R$} 
\end{cases}
\]
and all operations are performed over the field $K$.

It is not hard to check that if $(x_1,y_1)$ and $(x_2,y_2)$ are on the
curve, then so is $(x_3,y_3)$ and thus the above operation is closed
on $E$.  While not immediately clear, the points on the curve together
with the above operation form an abelian group (For a proof see
\cite{cha}).  It is clear from the definition of the group operation
that $\co$ is the identity element. If $P=(x,y) \in E$ it following
directly from equation \ref{eqnec1} that $R=(x,-y)$ is also a point on
the curve. Thus if $P=(x,y)$ then the inverse of $P$ is $(x,-y)$. Note
that the elliptic curve group operation is defined differently over
fields of characteristic $2$ or $3$.

A famous theorem by Hasse states that if $E$ is an elliptic curve
defined over $GF(p)$ then the number of points on $E$, denoted $\#E$,
is $p +1-t$ where $|t| \leq 2 \sqrt{p}$. This implies that the maximum
bit size of the order of a point is approximately the bit size of $p$.

A particular elliptic curve, $E$, is specified by giving the base
field, $K$, and the constants $a,b \in K$ from equation
$\ref{eqnec1}$. For our purposes the base field $K$ will always be
$GF(p)$ for some prime $p>3$.  In practice $a$ and $b$ are selected
such that the order of the group contains a large prime factor, $q$,
as this is necessary to make the discrete logarithm problem hard. For
simplicity we shall assume that $p$ and $q$ are approximately the same
bit size.

\subsection{Representing points on an elliptic curve} \label{rep_points}

Suppose we are given an elliptic curve, $E: y^2 = x^3+ax+b$, over the
field $GF(p)$ for some prime $p>3$. In order for a quantum computer to
calculate discrete logarithms over $E$ we like to have a unique
representation of the points on $E$.

If $P$ is a finite point on $E$ then $P=(x,y)$, where $x$ and $y$ are
integers modulo $p$. Thus any finite point can be represented by a
unique ordered pair $(x,y)$ with $x,y \in \{0,1, \dots ,p-1\}$.  Now
all that remains is to determine how $\co$ will be represented. As
will be discussed in section \ref{allow_err}, our implementation we
will not actually require a representation of $\co$.  However, if a
representation was required we could simply pick an ordered pair
$(x,y)$ which is not on the curve. For example, $(p,p)$ could be used
to represent $\co$ for any curve, while $(0,0)$ could be used for any
curve with $b\neq 0$.

\section{Our implementation of the quantum algorithm for discrete
logarithms over elliptic curves}

We consider an elliptic curve, $E$, over $GF(p)$, where $p$ is a large
prime. The base of the logarithm is a point $P \in E$ whose order is
another (large) prime $q$, thus $q P=\cal O$. We want to compute the
discrete logarithm, $d$, of another point $Q \in E$, thus $Q=d
P$. (Remember that we use additive notation for the group operation,
thus instead of a power of the base element, we have an integer
multiple.)

As discussed in section \ref{dlog}, we need to apply the following
transformation
\begin{equation*}
\frac{1}{2^n} \sum_{x=0}^{2^n-1} \sum_{y=0}^{2^n-1} \ket{x,y} \qquad \to
\quad \frac{1}{2^n}\sum_{x=0}^{2^n-1} \sum_{y=0}^{2^n-1} 
\ket{x,y,x P + y Q} 
\end{equation*}
Thus we need a method of computing (large) integer multiples of group
elements.  This can be done by the standard ``double and add
technique''. This is the same technique used for the modular
exponentiation in the factoring algorithm, although there the group is
written multiplicatively so it's called the square and multiply
technique.  To compute $x P +yQ$, first we repeatedly double the group
elements $P$ and $Q$, thus getting the multiples $P_i=2^i P$ and $Q_i
= 2^iQ$. We then add together the $P_i$ and $Q_i$ for which the
corresponding bits of $x$ and $y$ are 1, thus
\[
 x P +yQ = \sum_i x_i P_i + \sum_i y_i Q_i 
\]
where $x= \sum_i x_i 2^i$, $y= \sum_i y_i 2^i$, $P_i = 2^i P$ and $Q_i
= 2^i Q$.  The multiples $P_i$ and $Q_i$ can fortunately be
precomputed classically. Then to perform the above transformation,
start with the state $\sum_{x,y} \ket{x,y,\cal O}$. The third register
is called the ``accumulator'' register and is initialised with the
neutral element $\cal O$. Then we add the $P_i$ and $Q_i$ to this
register, conditioned on the corresponding bits of $x$ and $y$.

\subsection{Input registers can be eliminated} \label{g_shift}
 
Here we show that the input registers, $\ket{x,y}$, can actually be
shrunk to a single qubit, thus saving much space.  This is
accomplished by using the semiclassical quantum Fourier transform and
is completely analogous to what has been proposed for the factoring
algorithm \cite{cleve} (see also e.g. \cite{beau,zalka}).

Griffiths and Niu \cite{grif} have observed that the QFFT followed by
a measurement can be simplified. Actually it can be described as
simply measuring each qubit in an appropriate basis, whereby the basis
depends on the previous measurement results. (In accordance with
quantum computing orthodoxy, we can also say that before measuring the
qubit in the standard basis, we apply a unitary transformation which
depends on the previous measurement results.) Note that in the initial
state $ \frac{1}{2^n}\sum_{x=0, y=0}^{2^n-1,2^n-1}\ket{x,y,\cal O}$
the qubits in the $x$- and $y$-registers are actually
unentangled. Each qubits is in the state
$(\ket{0}+\ket{1})/\sqrt{2}$. Now we can see how these two registers
can be eliminated: We do $n$ steps. In step number $i$ we first
prepare a qubit in the state $(\ket{0}+\ket{1})/\sqrt{2}$, then use it
to control the addition of $P_i$ (or $Q_i$) and finally we measure the
control qubit according to the semiclassical QFFT. In this QFFT the
qubits have to be measured in reversed order, thus from highest
significance to lowest. Thus we will need to proceed from the $i=n-1$
step down to the $i=0$ step, but this is no problem.

In summary, we really only need the accumulator register. We are left
being required to carry out a number of steps whereby we add a fixed
(classically known) point $P_i$ (or $Q_i$) to a superposition of
points.  We are working in the cyclic group generated by $P$, thus the
effect of a fixed addition is to ``shift'' the discrete logarithm of
each element in the superposition by the same amount. For this reason
we shall refer to these additions of fixed classical points as ``group
shifts''. (That the group shifts are conditional on a qubit makes it
only insignificantly more difficult, as we will point out later.) Thus
we need unitary transformations $U_{P_i}$ and $U_{Q_i}$ which acts on
any basis state $\ket{S}$ representing a point on the elliptic curve,
as:
\[
 U_{P_i}: \quad \ket{S} ~\to~ \ket{S+P_i}  \hspace{.25in} \textrm{and} 
\hspace{.25in}  U_{Q_i}: \quad \ket{S} ~\to~ \ket{S+Q_i}
\]
As explained in section \ref{dlog} and appendix \ref{FT A}, it is
sufficient to do $n$ of these steps for $P$ and $n$ for $Q$, thus a
total of $2 n$, where $n \approx \log_2 q$.

\subsection{Simplifying the addition rule} \label{allow_err}

So we have already managed to decompose the discrete logarithm quantum
algorithm into a sequence of group shifts by constant classically
known elements. That is
$$ U_A : \ket{S} \to \ket{S+A} \qquad S,A \in E  ~ \textrm{ and } ~ 
A~\mbox{is fixed} 
$$
We propose to only use the addition formula for the ``generic'' case
(i.e. for $P+R$ where $P,R \neq \co$ and $P \neq \pm R$) for the group
operation, although it wouldn't be very costly to properly distinguish
the various cases. Still, it's not necessary. First note that the
constant group shifts $2^i P$ and $2^i Q$ are not equal to the neutral
element $\cal O$, because $P$ and $Q$ have order a large prime. (If a
group shift was $\cal O$, we would of course simply do nothing.)
Still, we are left with three problems. First, that a basis state in
the superposition may be the inverse of the group shift. Second, that
a basis state in the superposition may equal the group shift. Lastly,
that a basis state in the superposition may be $\co$.  We argue that
with a small modification these will only happen to a small fraction
of the superposition and thus the fidelity lost is negligible.
 
To ensure a uniformly small fidelity loss, we propose the following
modification at the beginning of the DLP algorithm: choose (uniformly)
at random an element $k \cdot P \neq \co$ in the group generated by
$P$. Then we initialise the accumulator register in the state $\ket{k
\cdot P}$, instead of $\ket{\cal O}$. This overall group shift is
irrelevant, as after the final QFFT it only affects the phases of the
basis states. Now on average in each group shift step we ``loose''
only a fraction of $1/q$ of the superposition by not properly adding
inverses of points and an equal amount for not correctly doubling
points.  Thus the total expected loss of fidelity from this error
during the $2n$ group shifts is $4n/q \approx 4 \log_{2} q / q$ and is
thus an exponentially small amount.  As the accumulator no longer
begins in the state $\ket{\co}$, the superposition $\ket{S}$ to which
$U_A$ will be applied can only contain $\ket{\co}$ if an inverse was
(correctly) added in the previous addition. Thus $\co$ being a basis
state in the superposition will not cause any further loss of
fidelity.
   
\subsection{Decomposition of the group shift}

The group shift is clearly reversible. A standard procedure might be
to do 
$\ket{S,0} \to \ket{S,S+A} \to \ket{0,S+A}$
where in the last step we would uncompute $S$ by running the addition
of $-A$ to $S+A$ backwards. Fortunately we can do better than this
generic technique. In terms of the coordinates of the points, the
group shift is:
$$ \ket{S}=\ket{(x,y)} \quad \to \quad \ket{S+A} =
\ket{(x,y)+(\alpha,\beta)}=\ket{(x',y')} $$ 
Recall that $x = \alpha$ if and only if $(x,y) = \pm A$ and that this
portion of the superposition can be lost (see section
\ref{allow_err}). Thus we use the following group operation formulas
(see eq. \ref{ECop}):
$$ \lambda=\frac{y-\beta}{x-\alpha}=-\frac{y'+\beta}{x'-\alpha} \qquad 
x'=\lambda^2-(x+\alpha) $$
The second expression for $\lambda$ is not difficult to obtain. It
will allow us to later uncompute $\lambda$ in terms of
$x',y'$. Actually, when computing $\lambda$ from $x,y$ we can directly
uncompute $y$, and similarly we can get $y'$ when uncomputing
$\lambda$:
$$ x,y \quad\leftrightarrow\quad x,\lambda \quad\leftrightarrow\quad x',\lambda
\quad\leftrightarrow\quad x',y' $$ 
Where a double-sided arrow ($\leftrightarrow$) indicates that we need
to do these operations reversibly, thus in each step we need also to
know how to go backward. Note that the decomposition of one large
reversible step into several smaller individually reversible ones, is
nice because it saves space, as any ``garbage'' can be uncomputed in
each small step. In more detail the sequence of operations is:
\begin{eqnarray} \label{eqnGS}
x,y ~\leftrightarrow~ x-\alpha,y-\beta &\leftrightarrow&
x-\alpha,\lambda=\frac{y-\beta}{x-\alpha} ~\leftrightarrow~ \\ \nonumber
&\leftrightarrow&
x'-\alpha,\lambda=-\frac{y'+\beta}{x'-\alpha} ~\leftrightarrow~
x'-\alpha,y'+\beta ~\leftrightarrow~ x',y'
\end{eqnarray}
where all the operations are done over $GF(p)$.  The second line is
essentially doing the operations of the first line in reverse. The
first and last steps are just modular additions of the constants $\pm
\alpha,-\beta$. They clearly need much less time (and also less
qubits) than the multiplications and divisions (see \cite{vbe}), so we
will ignore them when calculating the running times. The operation in
the middle essentially involves adding the square of $\lambda$ to the
first register. This operation, too, is relatively harmless. It uses
less ``work'' qubits than other operations and thus doesn't determine
the total number of qubits needed for the algorithm. Still, for time
complexity we have to count it as a modular multiplication (more about
this below). So a group shift requires two divisions, a multiplication
and a few additions/subtractions.

\subsubsection{Divisions of the form $x,y
\leftrightarrow x,y/x$} \label{g_less}

The remaining two operations are a division and multiplication where
one of the operands is uncomputed in the process. The division is of
the form $x,y \leftrightarrow x,y/x$, where $x \neq 0$.  (different
$x$ and $y$ than the last section!). The multiplication in
(\ref{eqnGS}) is simply the division run in the reverse direction.  We
further decompose the division into four reversible steps:
$$ x,y \quad\stackrel{E}{\leftrightarrow}\quad 
1/x,y \quad\stackrel{m}{\leftrightarrow}\quad 
1/x,y,y/x \quad\stackrel{E}{\leftrightarrow}\quad 
x,y,y/x \quad\stackrel{m}{\leftrightarrow}\quad 
x,0,y/x $$
Where the letters over the arrows are $m$ for ``multiplication'' and
$E$ for ``Euclid's algorithm'' for computing the multiplicative
inverse modulo $p$. The second $m$ is really a multiplication run
backwards to uncompute $y$.

\subsubsection{Modular multiplication of two ``quantum'' numbers}
\label{mm}

Before concentrating on Euclid's algorithm, let's look at the modular
multiplications of the form $x,y \leftrightarrow x,y,x \cdot y$. In
the quantum factoring algorithm the modular exponentiation is
decomposed into modular multiplications. But there one factor is a
fixed ``classical'' number. Still, the situation when we want to act
on superpositions of both factors, is not much worse. So we want to do
(explicitly writing mod $p$ for clarity):
$$ \ket{x,y} \quad \to \quad \ket{x,y,x\cdot y ~\mbox{mod}~ p} $$
We now decompose this into a sequence of modular additions and modular 
doublings:
$$ x\cdot y = \sum_{i=0}^{n-1} x_i 2^i y = x_0 y+2(x_1 y+2(x_2 y+2(x_3
y+ \dots))) \pmod{p}$$
So we do a series of the following operations on the third register:
$$ A \quad \leftrightarrow \quad 2 A  \quad \leftrightarrow \quad
2 A+x_i y \pmod{p} \qquad i=n-1 \dots 0$$

\subsubsection*{Modular doubling}

The modular doubling is a standard doubling (a left shift by one bit)
followed by a reduction mod $p$. Thereby we either subtract $p$ or
don't do anything. Whether we subtract $p$ has to be controlled by a
control qubit. At the end this control qubit can be uncomputed simply
by checking whether $2 A$ mod $p$ is even or odd (because $p$ is
odd). For the addition or subtraction of a fixed number like $p$ we
need $n$ carry qubits, which have to be uncomputed by essentially
running the addition backwards (but not undoing everything!). To do
the reduction mod $p$ we will now in any case subtract $p$, check
whether the result is negative, and depending on that, either only
uncompute the carry bits or undo the whole subtraction. In the end the
operation is only slightly more complicated than the addition of a
fixed number.

\subsubsection*{Modular addition}

The second step is a modular addition of the form $\ket{x,y} \to
\ket{x,x+y ~\mbox{mod}~ p}$. Again we first make a regular
addition. This is only slightly more complicated than the addition of
a fixed number (see e.g. \cite{zalka} pp. 7,8). Then, again, we either
subtract $p$ or not. To later uncompute the control bit which
controlled this, we have to compare $x$ and $x+y ~\mbox{mod}~ p$,
which essentially amounts to another addition. Thus overall we have
two additions.

So all together for the modular multiplication we have to do $n$
steps, each roughly consisting of 3 additions. So one multiplication
involves some $3 n$ additions.

\section{The Extended Euclidean Algorithm}

Suppose $A$ and $B$ are two positive integers. The well known
Euclidean algorithm can be used to find the greatest common divisor of
$A$ and $B$, denoted $\gcd(A,B)$.  The basis of the algorithm is the
simple fact that if $q$ is any integer then $\gcd(A,B) = \gcd(A, B -
qA)$. This implies the gcd doesn't change if we subtract a multiple of
the smaller number from the larger number.  Thus the larger number can
be replaced by its value modulo the smaller number without affecting
the gcd. Given $A$ and $B$ with $A\geq B$ this replacement can be
accomplished by calculating $q = \lfloor A/B \rfloor$ and replacing
$A$ by $A-qB$, where $\lfloor x \rfloor$ is the largest integer less
than or equal to $x$. The standard Euclidean algorithm repeats this
replacement until one of the two numbers becomes zero at which point
the other number is $\gcd(A,B)$. The table below illustrates the
Euclidean algorithm.

\vspace*{.1in}
\noindent \begin{tabular}{|c|l|c|c|l|} \hline
\multicolumn{2}{|c|}{$\gcd(A,B)$}               & & 
\multicolumn{2}{|c|}{$\gcd(1085,378)$}          \\ \hline
integers  & \multicolumn{1}{|c|}{quotient}       & &
 integers & \multicolumn{1}{|c|}{quotient}       \\ \hline 
$(A,B)$   & $q=\lfloor A/B \rfloor$             & &
$(1085,378)$    & $2=\lfloor 1085/378 \rfloor$  \\ \hline
$(A-qB,B)$& $q'=\lfloor \frac{B}{A-q B} \rfloor$& &
$(329,378)$     & $1=\lfloor 378/329 \rfloor$   \\ \hline
$(A-qB,B-q'(A-qB))$& $ q''=\dots$                & & 
$(329,49)$      & $6=\lfloor 329/49 \rfloor$    \\ \hline
$\cdot$   &  $\cdot$                            & & 
$(35,49)$       & $1=\lfloor 49/35 \rfloor$     \\ \hline
$\cdot$   &  $\cdot$                            & &
$(35,14)$       & $2=\lfloor 35/14 \rfloor$     \\ \hline       
$\cdot$   &  $\cdot$                            & & 
$(7,14)$        & $2=\lfloor 14/7 \rfloor$      \\ \hline       
$(\gcd(A,B),0)$   &                             & & 
$(7,0)$         &                               \\ \hline       
\end{tabular}

\vspace*{.1in}
\noindent It can be shown that the Euclidean algorithm will involve
$O(n)$ iterations (modular reduction steps) and has a running time of
$O(n^2)$ bit operations, where $n$ is the bit size of $A$ and $B$ (see
\cite{coh}).

Again suppose that $A$ and $B$ are two positive integers.  The
extended Euclidean algorithm can be used not only to find $\gcd(A,B)$
but also integers $k$ and $k'$ such that $kA+k'B = \gcd(A,B)$. This
follows from the fact that after each iteration of the Euclidean
algorithm the two integers are known integer linear combinations of
the previous two integers. This implies that the integers are always
integer linear combinations of $A$ and $B$.  The extended Euclidean
algorithm simply records the integer linear combinations of $A$ and
$B$ which yield the current pair of integers. Thus when the algorithm
terminates with $(\gcd(A,B),0)$ or $(0,\gcd(A,B))$ we will have an
integer linear combination of $A$ and $B$ which equals $\gcd(A,B)$.

Let us now turn our attention to finding $x^{-1} \pmod{p}$, for $x\neq
0$.  If the extended Euclidean algorithm is used to find integers $k$
and $k'$ such that $kx +k'p=1$ then $k \equiv x^{-1} \pmod{p}$. Note
that we are not interested in the coefficient $k'$ of $p$ in the
integer linear combination.  Thus we need only record the coefficient
of $x$ (and not $p$) in the extended Euclidean algorithm.

Hence to compute $x^{-1} \pmod{p}$ we will maintain two ordered pairs
$(a,A)$ and $(b,B)$, where $A$ and $B$ are as in the Euclidean
algorithm and $a$ and $b$ record the coefficients of $x$ in the
integer linear combinations.  We shall refer to these ordered pairs as
\textit{Euclidean pairs}.  (Note that $A$ and $B$ will equal $ax
\pmod{p}$ and $bx \pmod{p}$).  We begin the algorithm with
$(a,A)=(0,p)$ and $(b,B)=(1,x)$. In each iteration we replace either
$(a,A)$ or $(b,B)$. If $A \geq B$ then we replace $(a,A)$ with $(a-q b
, A-q B)$, where $q=\lfloor A/B \rfloor$. Otherwise $(b,B)$ is
replaced with $(b-qa , B-q A)$, where $q=\lfloor B/A \rfloor$.  The
algorithm terminates when one of the pairs is $(\pm p,0)$, in which
case the other pair will be $(x^{-1},1)$ or $(x^{-1}-p,1)$.  We
illustrate the algorithm in the following table.

\vspace*{.1in}
\noindent \begin{tabular}{|c|l|c|c|l|} \hline
\multicolumn{2}{|c|}{$x^{-1} \mod{p}$}                        & & 
\multicolumn{2}{|c|}{$96^{-1} \mod{257}$}                     \\ \hline
Euclidean pairs   & \multicolumn{1}{|c|}{quotient}             & &
Euclidean pairs   & \multicolumn{1}{|c|}{quotient}             \\ \hline 
$(0,p),(1,x)$ & $q=\lfloor p/x \rfloor$                       & & 
$(0,257),(1,96)$   & $2=\lfloor 257/96 \rfloor$          \\ \hline
$(-q,p-qx),(1,x)$ & $q'=\lfloor \frac{x}{p-qx} \rfloor$       & & 
$(-2,65), (1,96)$   & $1=\lfloor 96/65 \rfloor$           \\ \hline
\multicolumn{2}{|l|}{\small $(-q,p-qx),(1+q'q,x-q'(p-qx))$}   & &
 $(-2,65),(3,31)$   & $2=\lfloor 65/31 \rfloor$          \\ \hline
$\cdot$ &                                                     & &
 $(-8,3),(3,31)$    & $10=\lfloor 31/3 \rfloor$          \\ \hline
$\cdot$ &                                                     & &
 $(-8,3), (83,1)$    & $3=\lfloor 3/1 \rfloor$            \\ \hline
$ (-p,0),(x^{-1},1)$ &                                     & &
 $(-257,0),(83,1)$    &                                  \\ \hline
\end{tabular}
\vspace*{.1in}

\noindent Note that at termination the Euclidean pairs will either be
$(-p,0),(x^{-1},1)$ or $(x^{-1}-p,1),(p,0)$. In the later case we have
to add $p$ to $x^{-1}-p$ to get the standard representation.

\subsection{Stepwise reversibility}

A priori it's not clear whether an iteration of the extended Euclidean
algorithm is reversible. In particular it's not clear whether the
quotients $q$ will need to be stored or if they can be uncomputed. If
they need to be stored then this will constitute a considerable number
of ``garbage'' bits which could only be uncomputed (in the usual way)
once the whole inverse finding algorithm had finished. Fortunately it
turns out that each iteration of the algorithm is individually
reversible.

Concretely let's look at uncomputing the quotient $q$ after an
iteration which transformed $(a,A),(b,B)$ into $(a-qb,A-qB),(b,B)$. We
know that $A > B$ and $q= \lfloor A/B \rfloor$.  It is not hard to see
that $a$ and $b$ will never have the same sign and that $A >B$ if and
only if $|a| < |b|$. Therefore $\lfloor - \frac{a-qb}{b} \rfloor = q$.
Thus we see, that while $q$ is computed from the second components of
the original Euclidean pairs, it can be uncomputed from the first
components of the modified Euclidean pairs.

\subsection{Simple implementations with time $O(n^3)$ 
and $O(n^2 \log_{2} n)$}

While it is a relief that the extended Euclidean algorithm is
piecewise reversible, we are not at the end of our labours. Note that
the number of iterations (modular reductions) in the algorithm for
$x^{-1}$ mod $p$ depends on $x$ in an unpredictable way. This is a
problem because we want to apply this algorithm to a superposition of
many $x$'s. Still worse is, that even the individual iterations take
different times for different inputs $x$ when the algorithm is
implemented in an efficient way. Namely, the quotients $q$ tend to
be small, and we want to use algorithms in each iteration which
exploit this fact, since for small $q$ the steps can be made
faster. Only then does the extended Euclidean algorithm use time
bounded by $O(n^2)$.

Suppose that in each iteration of the algorithm we use full sized
divisions and multiplications, thus the ones which we would use if we
expected full sized $n$ bit numbers. These algorithms (e.g. the
modular multiplication described in section \ref{mm}) consist of a
fixed sequence of $O(n^2)$ gates and can work just as well on a
superposition of inputs. As there are $O(n)$ iterations, the extended
Euclidean algorithm would then use $O(n^3)$ gates.

\subsubsection{Using bounded divisions} \label{secBD}

The running time $O(n^3)$ can be improved by noting that large
quotients $q$ are very rare. Actually in a certain limit the
probability for the quotient to be $q_0$ or more, is given by $P(q \ge
q_0) = \log_2 (1+1/q_0) \approx 1/(q_0 \ln 2)$ (see e.g. \cite{knuth}
Vol. 2, section 4.5.3). If we use an algorithm that works for all
quotients with less than, say, $3 \log_{2} n$ bits, then the
probability of error per iteration will be $\approx 1/n^3$. Or, if
acting on a large superposition, this will be the fidelity
loss. Because in the whole discrete logarithm algorithm we have
$O(n^2)$ such iterations ($O(n)$ iterations for each of the $O(n)$
group shifts), the overall fidelity loss will only be of order
$O(1/n)$. Still, even with these bounded divisions the overall
complexity of the extended Euclidean algorithm would be $O(n^2\log_{2}
n)$.

We would like to obtain a running time of $O(n^2)$, which would lead
to an $O(n^3)$ discrete logarithm algorithm.  Our proposed
implementation of the extended Euclidean algorithm attains this
$O(n^2)$ running time. Our implementation is not only faster
asymptotically, but also for the sizes $n$ of interest, although only
by a factor of 2 to 3.

\subsection{Our proposed implementation}

We have investigated various efficient implementations of the extended
Euclidean algorithm. Fortunately, the one presented here is one of the
simpler ones.  To get an $O(n^2)$ algorithm, we will not require all
the basis states in the superposition to go through the iterations of
Euclid's algorithm synchronously. Rather we will allow the computation
for each basis state to proceed at its own pace. Thus at a given time,
one computation may be in the course of the $10$-th iteration, while
another one is still in the $7$-th iteration. Later the second
computation may again overtake the first one.

The basic observation of our implementation is that it consists of
only five different operations, most of which are essentially
additions and subtractions. Thus each of the many ``quantum-parallel''
computations (thus each basis state) can store in a few flag bits
which one of these five operations it needs.  The implementation can
then simply repeatedly cycle through the five operations one after
the other, each one conditioned on the flag qubits. Thus as each
operation is applied only those basis states which require it will be
affected.  For each cycle through the five operations the flag bits of
a given computation will often only allow one operation to be applied
to the computation. Therefore we loose a factor of somewhat less than
five in speed relative to a (reversible) classical implementation.

\subsubsection{Desynchronising the parallel computations}
\label{desync}

Let us first explain in more detail the general approach to
desynchronising the ``quantum-parallel'' computations. Suppose, for
example, that there are only three possible (reversible) operations
$o_1, o_2$ and $o_3$ in a computation. Suppose further that each
computation consists of a series of $o_1$'s then $o_2$'s, $o_3$'s and
so on cyclicly. E.g. we would like to apply the following sequence of
operations to two different basis state:
\begin{eqnarray*}
\dots o_2 o_2 o_2 ~~ o_1 ~~  o_3 o_3 ~~ o_2 ~~ o_1 o_1 o_1 o_1 ~\ket{x} && \\
\dots o_2 ~~ o_1 o_1 o_1 ~~ o_3 ~~ o_2 o_2 o_2 o_2 ~~ o_1 ~\ket{x'} &&
\end{eqnarray*}
Clearly there must be a way for the computation to tell when a series
of $o_i$'s is finished and the next one should begin. But because we
want to do this reversibly, there must also be a way to tell that an
$o_i$ is the first in a series. Say we include in each $o_i$ a
sequence of gates which flips a flag qubit $f$ if $o_i$ is the first
in a sequence and another mechanism that flips it if $o_i$ is the last
in a sequence. (If there is a single $o_i$ in a sequence, and thus
$o_i$ is both the first and the last $o_i$, then $f$ is flipped
twice.)

We will also make use of a small control register $c$ to record which
operation should be applied to a given basis state.  Thus we have a
triple $~x,f,c~$ where $x$ stands for the actual data. We initialise
both $f$ and $c$ to $1$ to signify that the first operation will be
the first of a series of $o_1$ operations.  The physical quantum-gate
sequence which we apply to the quantum computer is:
$$ \dots ac~ o'_1~~~ac~ o'_3~ac~ o'_2~ac~  o'_1~~~ac~ o'_3~ac~ 
 o'_2~ac~  o'_1~\ket{QC} $$
Where the $o'_i$ are the $o_i$ conditioned on $i=c$ and $ac$ stands
for ``advance counter''. These operations act as follows on the
triple:
\begin{eqnarray*}
o'_i : && \mbox{if}~~ i=c~: \qquad x,f,c \quad\leftrightarrow\quad
 o_i(x),f \oplus first \oplus last,c \\
ac : && x,f,c \quad\leftrightarrow\quad x,f,(c+f) ~\mbox{mod}~ 3
\end{eqnarray*}
Where $o'_i$ doesn't do anything if $i \not = c$, $\oplus$ means XOR
and $(c+f) ~\mbox{mod}~ 3$ is taken from $\{1,2,3\}$.  In the middle
of a sequence of $o_i$'s the flag $f$ is $0$ and so the counter
doesn't advance. The last $o_i$ in a series of $o_i$'s will set $f=1$
and thus the counter is advanced in the next $ac$ step. Then the first
operation of the next series resets $f$ to $0$, so that this series
can progress.

\subsubsection{Applying this to the extended Euclidean algorithm}

Back to our implementation of the extended Euclidean algorithm. For
reasons which will be discussed below, in our implementation we always
store the Euclidean pair with the larger second coordinate first.
Thus one (reversible) iteration of the algorithm is:
$$ (a,A),(b,B) \quad\leftrightarrow\quad (b,B),(a-q b,A-q B) \qquad 
\mbox{where} \quad  q=\lfloor A/B \rfloor =\lfloor -\frac{a-q b}{b}
\rfloor $$
This will be decomposed into the following three individually reversible
steps: 
\begin{equation} \label{SWAP}
A,B ~\leftrightarrow~ A-q B,B,q \quad \qquad a,b,q
~\leftrightarrow~ a-q b,b \quad \qquad \mbox{and} \qquad \mbox{SWAP} 
\end{equation}
where by ``SWAP'' we mean the switching of the two Euclidean pairs.
Note that it appears that all the bits of $q$ must be calculated
before they can be uncompute. This would mean that the computation can
not be decomposed further into smaller reversible steps.  We now
concentrate on the first of these operations, which starts with a pair
$A,B$ of positive integers where $A > B$. Actually, since $|a-q b| >
|b|$, the second operation $a,b,q\leftrightarrow a-q b,b$ can be
viewed as the same operation run backwards. The fact that $a$ and $b$
can be negative (actually they have opposite sign) is only a minor
complication.

So we want to do the division $A,B\leftrightarrow A-q B,B,q$ in a way
which takes less time for small $q$, namely we want to do only around
$\log_{2} q$ subtractions. What we do is essentially grade school long
division in base $2$. First we check how often we have to double $B$
to get a number larger than $A$, and then we compute the bits of $q$
from highest significance to lowest. In the first phase (operation
$1$) we begin with $i=0$ and do a series of operations of the form
$$ A,B,i \quad\leftrightarrow\quad A,B,i+1 $$
As explained in section \ref{desync}, we need to flip a flag bit $f$
at the beginning and at the end of this series. The beginning is
easily recognised as $i=0$. The end is marked by $2^i B > A$. Thus
testing for the last step essentially involves doing a subtraction in
each step.

In the second phase (operation $2$) we then diminish $i$ by $1$ in each step:
$$ A-q' B,~B,~i+1,~q' \quad\leftrightarrow\quad  
A-(q'+2^i q_i) B,~B,~i,~q'+2^i q_i $$
where $q'=2^{i+1} q_{i+1}+2^{i+2} q_{i+2} + \dots$ is the higher order
bits of $q$. The new bit $q_i$ is calculated by trying to subtract
$2^i B$ from the first register. This is easily done by subtracting
and, if the result is negative ($q_i=0$), undoing the entire
subtraction in the carry uncomputing phase. The first operation in the
second phase is recognised by checking $q'=0$ and the last by checking
$i=0$. Note that when incrementing and decrementing $i$ we might
actually want to shift the bits of $B$ to the left (resp. right), as
then the gates in the subtractions can be done between neighbouring
qubits.

The third and fourth phases perform $a,b,q \leftrightarrow a-q b,b$
and are essentially the reverses of phases two and one
respectively. Thus operations three and four are
\[
 a-qb +(q'+2^{i}q_i)b,~b,~i,~q'+2^i q_i \quad\leftrightarrow\quad  
a-qb +q'b,~b,~i+1,~q'
\]
and
\[
 a-qb,b,i \quad\leftrightarrow\quad a-qb,b,i-1
\]
where $q'$ and $q_i$ are as in phase two. The first and last operation
conditions of phase three are $i=0$ and $q'=0$. While the first and
last operation conditions of phase four are $|a-qb| < 2^{i+1}|b|$ and
$i=0$.  (These conditions are essentially the last and first operation
conditions for phases two and one respectively.)
 
Finally we also need to do the SWAP operation where we switch the two
Euclidean pairs so that again their second components are in the
order: larger to the left, smaller to the right. If we didn't do that,
we would have to double the set of operations above, to take care of
the case when the roles of the pairs are switched. The SWAP of course
simply switches the registers qubit by qubit (although this has to be
done conditional on the control register $c$). As every sequence of
SWAP operations will be of length one, the flag bit $f$ can be left
untouched.

\subsubsection{How many steps are necessary?} \label{secNumSteps}

With the SWAP we have a sequence of five operations which we
repeatedly apply one after the other to the quantum computer. So the
question is: How many times do we need to cycle through the five
operations?  Each iteration of Euclid's algorithm ends with a SWAP and
thus requires an integer number of cycles through the five operations.
Also note, that in each iteration the length of the sequence of
operations $o_i$ is the same for all four phases. Thus for one
iteration the following operations might actually be applied to one
computation:
$$ \mbox{SWAP}~o_4~o_4~o_4~~o_3~o_3~o_3~~o_2~o_2~o_2~~o_1~o_1~o_1 \ket{x} 
$$
The length $z$ of the sequences (here 3) is the bit length of the
quotient $q$ in the iteration (i.e. $z=\lfloor \log_{2} q \rfloor
+1$). If each operation is done only once (so $z=1$), everything will
be finished with one cycle through the five operations. In general the
number of cycles for the iteration will be $4(z-1)+1$.

Let $r$ be the number of iterations in a running of Euclid's Algorithm
on $p,x$, let $q_1,q_2, \dots ,q_r$ be the quotients in each iteration
and let $t$ be the total number of cycles required.  Then
\[
t =\sum_{i=1}^{r}(4\lfloor \log_{2}(q_i) \rfloor+1)  
  = r + 4\sum_{i=1}^{r}\lfloor \log_{2}(q_i) \rfloor
\]
For $p>2$ a bound on $t$ is $4.5 \log_{2}(p)$ (see appendix
\ref{AppB}). Thus we can bound the number of cycles by $4.5n$.

\subsubsection{The quantum halting problem: a little bit of garbage} 
\label{secgarbage}

Actually, because the inverse computation for each basis state has to
be reversible it can't simply halt when $B=0$. Otherwise, when doing
things backward, we wouldn't know when to start uncomputing.  This has
been called the ``quantum halting problem'', although it seems to have
little in common with the actual classical (undecidable) halting
problem. Anyway, instead of simply halting, a computation will have to
increment a small ($\log_{2}{4.5n}$ bit) counter for each cycle after
Euclid's algorithm has finished. Thus once per cycle we will check if
$B=0$ to determine if the completion counter needs to be incremented.
This means that at the end of Euclid's algorithm we will actually have
a little ``garbage'' besides $x^{-1}~\mbox{mod}~p$. Also, at least one
other bit of garbage will be necessary because, as mentioned earlier,
half the time we get $x^{-1}-p$ instead of $x^{-1}$ itself. Note that
in our computation of $x,y \leftrightarrow x,y/x$ we use Euclid's
algorithm twice, once for $x \leftrightarrow x^{-1}$ and once for the
converse (see section \ref{g_less}).  Thus we can simply leave the
garbage from the first time around and then run the whole algorithm
backwards.

\subsubsection{Saving space: Bounded divisions and register sharing}
 \label{secSaveSpace}

To solve the DLP our algorithm will need to run the Euclidean
algorithm $8 n$ times. Each running of the Euclidean algorithm will
require at most $1.5n$ iterations (see \cite{sha}).  Thus the DLP
algorithm will require at most $12n^2$ Euclidean iterations. As
mentioned in section \ref{secBD}, the probability for the quotient,
$q$, in a Euclidean iteration to be $c \log_{2}{n}$ or more, is
$\approx 1/n^c$. Thus by bounding $q$ to $3 \log_{2}{n}$-bits (instead
of $n$ bits) the total loss of fidelity will be at most $12/n$.

Over the course of Euclid's algorithm, the first number $a$ in the
Euclidean pair $(a,A)$ gets larger (in absolute value), while $A$ gets
smaller. Actually the absolute value of their product is at most $p$:
At any given time, we store the two parentheses $(a,A)$ and
$(b,B)$. It is easy to check that $|b A-a B|$ remains constant and
equals $p$ during the algorithm ($b A-a B$ simply changes sign from
one iteration to the next and the initial values are $(0,p)$ and
$(1,x)$).  Now $p=|b A-a B| \ge |b A| \ge |a A|$, where we used that
$a$ and $b$ have opposite sign and $|a| < |b|$. So we see that $a$ and
$A$ could actually share one $n$-bit register. Similarly, since $|bB|
\leq |bA| \leq p$, it follows that $b$ and $B$ could also share an
$n$-bit register.

The problem is, that in the different ``quantum parallel''
computations, the boundary between the bits of $a$ and those of $A$
(or $b$ and $B$) could be in different places. It will be shown in
section \ref{improve} that the average number of cycles required is
approximately $3.5n$. Thus on average after $r$ cycles we would expect
$A$ and $B$ to have size $n -r/3.5$ and the size of $a$ and $b$ to be
$r/3.5$. We shall define the ``size perturbation'' of a running of the
extended Euclidean algorithm as the maximum number of bits any of
$A,B,a$ or $b$ reach above their expected sizes. Table \ref{tableDiff}
gives some statistics on size perturbations for various values of
$n$. For each value of $n$ in the table, the size perturbations were
calculated for one million runnings of Euclid's algorithm ($1000$
random inverses for each of $1000$ random primes). From the table we
see that the mean of the size perturbations ranges from
$1.134\sqrt{n}$ for $n=110$ to $1.069\sqrt{n}$ for $n=512$ and over
all $6$ million calculations was never over $2\sqrt{n}$. By analyzing
the distributions of the size perturbations it was seen that for $n
\in [110,512]$ the distributions are close to normal with the given
means and standard deviations.

Thus one method of register sharing between $a, A, b$ and $B$ would be
to take the size of the registers to be their expected values plus
$2\sqrt{n}$. In this case the four registers could be stored in $2n+
8\sqrt{n}$ qubits (instead of $4n$ qubits). Note that $a,A,b$ and $B$
are never larger than $p$, thus when implementing the register sharing
one would actually use the minimum of $n$ and the expected value plus
$2\sqrt{n}$. As the amount of extra qubits added to the expected sizes
of the buffers was only found experimentally, we shall carry through
the analysis of the algorithm both with and without register sharing.
   
\begin{table}[tbp] 
\begin{center}
\begin{tabular}{|c|c|c|c|c|} \hline
 $n$ & Mean Size    & Standard & Maximum Size\\
     & Perturbation & Deviation & Perturbation \\ \hline
$110$ & $11.90$     & $1.589$   & $18$ \\ \hline
$163$ & $14.13$     & $1.878$   & $24$ \\ \hline
$224$ & $16.35$     & $2.115$   & $25$ \\ \hline
$256$ & $17.33$     & $2.171$   & $25$ \\ \hline
$384$ & $21.02$     & $2.600$   & $31$ \\ \hline
$512$ & $24.20$     & $3.084$   & $38$ \\ \hline
\end{tabular}
\caption{\label{tableDiff} Size perturbations during Euclid's Algorithm}
\end{center}
\end{table}

\subsection{Analysis of the  Euclidean algorithm implementation}

The most basic operations, namely additions and subtractions, are in
the end conditioned on several qubits, which seems to complicate
things a lot. But before e.g. doing an addition we can simply compute
the AND of these control qubits, put it into an auxiliary qubit, and
use this single qubit to control the addition. Thus the basic
operations will essentially be (singly) controlled additions and
subtractions, as for the factoring algorithm. Detailed networks for
this can e.g. be found in \cite{zalka}.

\subsubsection{Running time:  $O(n^2)$}  \label{secEAruntime}

Let us now analyze the running time of our implementation of the
extended Euclidean algorithm. The algorithm consists of $4.5n$
operation cycles. During each of these cycles the halting register
needs to be handled and each of the five operations needs to be
applied.

Handling the halting register requires checking if $B=0$ and
incrementing the $\log_{2}{4.5n}$ bit register accordingly.  The
following table summarises the operations required in the first four
operations of a cycle.\\
\begin{center}
\noindent \vspace*{.1in}\begin{tabular}{|c|c|c|c|} \hline
  & Main Operation & First Check & Last Check \\ \hline 
$1$ & $z$-bit ADD  & $z$-bit ZERO& $w$-bit SUB \\ \hline
$2$   & $w$-bit SUB, $z$-bit SUB  & $(3\log_{2}{n})$-bit ZERO & 
$z$- bit ZERO \\ \hline 
$3$   & $w$-bit ADD, $z$-bit ADD & $z$- bit ZERO & 
$(3\log_{2}{n})$-bit ZERO \\ \hline
$4$ & $z$-bit SUB  & $w$-bit SUB & $z$-bit ZERO  \\ \hline 
\end{tabular} \\
\end{center}
Where $z=\log_{2}(3 \log_{2} n)$ is the size of the register for $i$,
$w$ represents the bit size of the registers for $a,A,b$ and $B$ ($w
\leq n$ and depends on whether or not register sharing is being used),
ZERO means a compare to zero and the $w$-bit operations are applied to
two quantum registers. Lastly, the fifth operation of the cycle, SWAP,
swaps two registers of size at most $2n$.

Therefore each of the $4.5n$ cycles requires $4$ $w$-bit
additions/subtractions, a SWAP and a $w$-bit compare to zero, all of
which are $O(n)$. The running time of the $w$-bit operations dominate
the algorithm and lead to a running time of $O(n^2)$.

\subsubsection{Space: $O(n)$} \label{space}

Let us now determine the number of qubits required for our
implementation of the extended Euclidean algorithm.  The largest
storage requirement is for the two Euclidean pairs $(a,A)$ and
$(b,B)$, which as discussed in section \ref{secSaveSpace}, can be
either $4n$ or $2n+8\sqrt{n}$ bits depending on whether or not
register sharing is used.  The next largest requirement is the $n$
bits needed for the carry register during the additions and
subtractions. The quotient $q$ will require $(3\log_{2}{n})$ bits (see
\ref{secSaveSpace}). The halting counter, $h$, will be of size
$\log_{2}{4.5n}$, however since $h$ and $q$ are never required at the
same time they can share a register.  The $i$ register needs to be
able to hold the bit size of the maximum allowed quotient ($3 \log_{2}
n$) and thus easily fits in a $\log_{2}{n}$ register.  Lastly the
algorithm requires a small fixed number ($< 10$) of bits for the flag
$f$, the control register $c$ and any other control bits. Thus we see
that the algorithm requires approximately $5n+ 4\log_{2}{n}+\epsilon$
or $3n+ 8\sqrt{n} +4\log_{2}{n}+\epsilon$ bits depending of whether
register sharing is used. In either case we see that the space
requirement is $O(n)$.
 
\subsubsection{Possible improvements and alternative approaches} 
\label{improve}

Here we list a few possible improvements and alternatives to our
approach.  It might also be that there are standard techniques, which
we are not aware of, for finding (short) acyclic reversible circuits.

\subsubsection*{Reducing the number of cycles}

While $4.5n$ is a limit on the maximum number of cycles required in
the Euclidean algorithm, there are very few inputs for which the
algorithm actually approaches this bound. For a prime $p$ let $L_q(p)$
be the number of times $q$ occurs as a quotient when the Euclidean
algorithm is run on $p,x$ for all $x$ satisfying $1 <x <p$. In
\cite{Heil} it was shown that
\[
L_q(p) = \frac{12(p-1)}{\pi^2} \ln \biggl(\frac{(q+1)^{2}}{(q+1)^{2}-1}\biggr)
\ln(p) + O(p(1+1/p)^{3}) 
\]
Using this fact, it can be shown that the total number of cycles
 required for finding $x^{-1}$ for all $1< x <p$ is
\begin{eqnarray*}
 \sum_{q=1}^{p-1} L_q(p) \bigl(4\lfloor \log_{2}(q) \rfloor +1) \approx 
(p-1) 3.5 \log_{2}(p) 
\end{eqnarray*} 
Thus the average number of cycles is approximately $3.5n$. Experiments
conducted seem to show that the distribution of required cycles is
close to normal with a standard deviation of around $\sqrt{n}$.  Thus
if we run the quantum computer a few standard deviations beyond the
average number of cycles, nearly all computations will have halted
making the loss of fidelity minimal. While this still leads to a
$O(n^3)$ DLP algorithm, the constant involved will have decreased.

\subsubsection*{Reducing the number of carry qubits}

Actually the number of carry qubits can be reduced by ``chopping''
e.g. an $n$-qubit addition into several pieces and have the carry
qubits not go much beyond one piece at a time. (Thereby we sacrifice
some fidelity, see e.g. \cite{zalka}.) This procedure takes somewhat
more time than a standard addition, but it may well make sense to
reduce the number of carry qubits (currently $n$) by a factor of 2 or
3.

\subsubsection*{Store length of numbers separately}

Here the idea is to also store the bit lengths of the numbers in
$(a,A)$ and $(b,B)$. In the divisions $A/B$ etc. the size of $q$ could
be determined by one comparison. Also the register sharing might be
easier, allowing for fewer than the current $8\sqrt{n}$ extra qubits.
Another possibility might be to synchronise the quantum parallel
computations by the lengths of the numbers. Then we would e.g. even
classically know the size of $A$.

\subsubsection*{More classical pre-computation for $GF(p)$}

As mentioned earlier, we can assume that classical computation is much
cheaper than quantum computation. Thus it might be reasonable to
classically pre-compute and store many values specific to $GF(p)$, if
these values would help to make the quantum implementation of Euclid's
algorithm easier. Unfortunately we haven't found any way of doing
this.

\subsubsection*{A quantum solution to arithmetic in $GF(p)$}

With an (approximate) quantum Fourier transform of size p, addition
modulo $p$ in a way becomes simpler \cite{draper}. It would be nice to
find such a ``quantum solution'' to both, addition {\it and}
multiplication. But to us it seems unlikely that this is possible.

\subsubsection*{Binary extended Euclidean algorithm}

This is a variant of Euclid's algorithm (see e.g. \cite{knuth},
Vol. 2, p. 338) which only uses additions, subtractions and bit shifts
(divisions by 2). Basically one can subtract $(b,B)$ from $(a,A)$, but
one also divides a parenthesis by 2 till the second component is
odd. We haven't managed to see that this algorithm is piecewise
reversible. Still, even if it isn't, it may be enough to keep
relatively little garbage around to make it reversible (Our
implementation of this algorithm used $7n+\epsilon$ qubits and had a
running time of $O(n^2)$).

\section{Results and a comparison with factoring}

\subsection{Total time for the DLP algorithm}

Let's collect the total number of quantum gates for the whole discrete
logarithm algorithm. Remember that the success probability of the
algorithm is close to 1 (appendix \ref{FT A}), thus we assume that we
have to do only a single run. We will not actually go to the lowest
level and count the number of gates, but rather the number of
($n$-bit) additions.

In table \ref{tableOps}, we decompose each part of the DLP algorithm
into its subroutines, plus things that can be done directly (to the
right). At the top are the $2n$ group shifts by a fixed (classical)
elliptic curve point (section \ref{g_shift}), $n$ for $x \cdot P$ and
$n$ for $y \cdot Q$.  Each group shift is decomposed into 2 divisions
(section \ref{g_less}), a multiplication to square $\lambda$, and a
few modular additions.  Multiplications and additions here are
understood to be modulo $p$ (section \ref{mm}).

\begin{table}[h]
\[  
\begin{array}{l}
\underbrace{\mbox{$2n$ group shifts}} 
~~(\mbox{e.g.}~~\ket{A} \to \ket{A+2^i \cdot P})
\vspace{5pt} \\
\qquad \quad \downarrow \mbox{each} \vspace{5pt} \\
\qquad \underbrace{\mbox{2 divisions}}~~ + 
\underbrace{\mbox{1 multiplication}~ 
(\mbox{for squaring $\lambda$})}_{\qquad \quad \mbox{each $3 n$ additions}}
~~ + \mbox{5 additions} 
 \\
\qquad \qquad \quad \downarrow \mbox{each} \vspace{5pt} \\ 
\qquad \qquad \underbrace{\mbox{2 Euclid's}}  ~~+~~ 
\underbrace{\mbox{2 multiplications}}_
{\quad \qquad \mbox{each $3 n$ additions}} \\ 
\qquad \qquad \qquad \quad  \downarrow \mbox{each}\vspace{5pt} \\ 
\qquad \qquad \qquad 
\underbrace{\mbox{$4.5 n$ cycles}} \vspace{5pt}
\\
\qquad \qquad \qquad \qquad \quad 
\downarrow \mbox{each}\vspace{5pt} \\ 
\qquad \qquad \qquad \qquad \underbrace{\mbox{5 operations}} ~~+~~
\mbox{halting counter}
\vspace{5pt}\\ 
\qquad \qquad \qquad \qquad \qquad 
\quad \downarrow \mbox{each}\vspace{5pt} \\ 
\qquad \qquad \qquad \qquad
\underbrace{\mbox{1 (short) addition}}_
{\mbox{average $\approx n/2+2\sqrt{n}$ bits}} +~~ 
\underbrace{\mbox{flag + counter operations}}_{\mbox{on $ \leq 3\log_2 n$ 
bit registers}}
\end{array}
\]
\caption{\label{tableOps} DLP Algorithm Operations } 
\end{table}

As discussed in section \ref{secEAruntime}, a running of Euclid's
algorithm requires $4.5n$ cycles through the five operations. If $w$
represents the sizes of the $a,A,b$ and $B$ registers then each of
these cycles requires a $w$-bit compare to zero (for the halting
register), $4$ $w$-bit additions/subtractions, a register swap and
some operations on $3\log_{2}{n}$ and $\log_2 (3\log_2 n)$ bit
registers. For our analysis we shall assume that all these operations
together are equivalent to $5$ $w$-bit additions (This is reasonable
since the SWAP and compare to zero operations are quite easy compared
to additions, see e.g. \cite{beau} fig. 10).  After the first running
of Euclid's algorithm we have found the inverse, but still have some
garbage in the halting register. We saw in section \ref{secgarbage}
that the second running of Euclid's algorithm will actually be the
reverse of the above operations. Thus the running time of the two
instances of Euclid's algorithm will be double the above operations.

To get a nice comparison to the factoring algorithm we need to know
how many classical-quantum additions are required (since the factoring
algorithm uses additions in which one summand is known classically
\cite{zalka}). In order to do this we assume that a quantum-quantum
addition is a factor of 1.7 times more difficult than a
classical-quantum addition (we estimated this ratio from the networks
in \cite{zalka}). When register sharing is used, the sizes of the
$a,A,b$ and $B$ registers change linearly between $2\sqrt{n}$ and
$n+2\sqrt{n}$. This implies that on average $w=n/2 +2\sqrt{n}$.  This
gives a total running time of
\[
 T = 2n \bigl[5+ 3n+ 2[6n+2(4.5(5n))]\bigr] \cdot 1.7  
\approx 360 n^2
\]
$n$-bit additions with no register sharing and 
\[
T = 2n \bigl[5+ 3n+ 2[6n+2(4.5\cdot 5(n/2+2\sqrt{n}))]\bigr] \cdot 1.7 
\approx 205 n^2 + 615n^{3/2} 
\] 
$n$-bit additions with register sharing.

As a classical-quantum addition is $O(n)$ this implies that the
discrete logarithm algorithm is $O(n^3)$. Assume a running time of
$k\cdot n$ for an $n$-bit classical-quantum addition. Then the
discrete logarithm algorithm has a running time of approximately $360
k n^3$ compared to only about $4 k n^3$ for factoring, but the larger
$n$ needed for classical intractability more than compensates for this
(see section \ref{secDLPvFact}).

\subsection{Total number of qubits (roughly $6n$)} \label{sec:num:bits}

For the number of qubits necessary for the discrete logarithm
algorithm, what counts is the operations during which the most qubits
are needed.  Clearly this is during the extended Euclidean algorithm,
and not e.g. in the course of a modular multiplication.

In fact, the maximum qubit requirement will occur in the second call
to the Euclidean algorithm within each division (see section
\ref{g_less}).  Here we require two $n$-bit registers plus a register
on which to carry out the Euclidean algorithm (see table
\ref{tableBits}).  Thus the DLP algorithm requires either $f(n) = 7n+
4\log_{2}{n}+\epsilon$ or $f'(n) = 5n+ 8\sqrt{n}
+4\log_{2}{n}+\epsilon$ bits depending of whether register sharing is
used (see section \ref{space}).  Therefore the DLP algorithm, like the
extended Euclid algorithm, uses space $O(n)$.
\begin{table}[h] 
$
\begin{array}{l}
\underbrace{\mbox{division}} \\
\qquad \downarrow \vspace{5pt} \\
\qquad 2n + \underbrace{\mbox{Euclid}} \\
\qquad \qquad \qquad \downarrow \vspace{5pt} \\ 
\qquad \qquad \underbrace{(a,A)+ (b,B)}_{\mbox{$2n+8\sqrt{n}$}} ~+~ 
\underbrace{\mbox{carry qubits}}_{\mbox{$n$ qubits}} ~+~ 
\underbrace{q ~+~ i}_{\mbox{$4\log_2{n}$ qubits}} 
~+~\underbrace{\mbox{minor stuff}}_{\mbox{$< 10$ qubits}}
\end{array}
$
\caption{\label{tableBits} Maximum Bit Requirement With Register Sharing} 
\end{table}

\subsection{Comparison with the quantum factoring algorithm}
\label{secDLPvFact}

One of the main points of this paper is that the computational
``quantum advantage'' is larger for elliptic curve discrete logarithms
than for the better known integer factoring problem. With our proposed
implementation we have in particular achieved similar space and time
requirements. Namely the number of qubits needed is also of $O(n)$ and
the number of gates (time) of order $O(n^3)$, although in both cases
the coefficient is larger. Note that the input size $n$ is also the
key size for RSA resp. ECC public key cryptography. Because the best
known classical algorithms for breaking ECC scale worse with $n$ than
those for breaking RSA, ECC keys with the same computational security
level are shorter. Below is a table with such key sizes of comparable
security (see e.g. \cite{ecc_rsa}). The column to the right roughly
indicated the classical computing resources necessary in multiples of
$C$, where $C$ is what's barely possible today (see. e.g. the RSA
challenges \cite{RSAc} or the Certicom challenges
\cite{Certicomc}). Breaking the keys of the last line seems to be
beyond any conceivable classical computation, at least if the
presently used algorithms can't be improved.
 
\noindent \vspace*{.1in}\begin{tabular}{|c|c|c||c|c|c||c|} \hline
\multicolumn{3}{|c||}{\mbox{Factoring algorithm (RSA)}} &
\multicolumn{3}{|c||}{\mbox{EC discrete logarithm (ECC)}} &
\multicolumn{1}{c|}{\mbox{classical}} \\ \hline
$n$    & {\small$\approx$ \# qubits}& time  & $n$  &  {\small$\approx$ \# 
qubits} & time & time  \\ \hline
       & $2n$    & $4n^3$  &  & {\small $f'(n)$ $(f(n))$}     & 
$360n^3$    &       \\ \hline \hline
$512$  & $1024$  & $0.54 \cdot 10^9$   &$110$ & $700$ $(800)$   & 
$0.5 \cdot 10^9$ & $C$ \\ \hline
$1024$ & $2048$  & $4.3 \cdot 10^9$    &$163$ & $1000$ $(1200)$  & 
$1.6 \cdot 10^9$ & $C \cdot 10^8$ \\ \hline
$2048$ & $4096$  & $34 \cdot 10^9$     &$224$ & $1300$ $(1600)$ & 
$4.0 \cdot 10^9$ & $C \cdot 10^{17}$ \\ \hline
$3072$ & $6144$  & $120 \cdot 10^9$    &$256$ & $1500$ $(1800)$ & 
$6.0 \cdot 10^9$ & $C \cdot 10^{22}$ \\ \hline
$15360$& $30720$ & $1.5 \cdot 10^{13}$&$512$ & $2800$ $(3600)$ & 
$50  \cdot 10^9$ & $C \cdot 10^{60}$ \\ \hline
\end{tabular} \\
Where $f(n)$ and $f'(n)$ are as in section \ref{sec:num:bits} with
$\epsilon=10$. The time for the quantum algorithms is listed in units
of ``1-qubit additions'', thus the number of quantum gates in an
addition network per length of the registers involved. This number is
about 9 quantum gates, 3 of which are the (harder to implement)
Toffoli gates (see e.g. \cite{zalka}). Also it seems very probable
that for large scale quantum computation error correction or full
fault tolerant quantum computation techniques are necessary. Then each
of our ``logical'' qubits has to be encoded into several physical
qubits (possibly dozens) and the ``logical'' quantum gates will
consist of many physical ones. Of course this is true for both quantum
algorithms and so shouldn't affect the above comparison. The same is
true for residual noise (on the logical qubits) which will decrease
the success probability of the algorithms. The quantum factoring
algorithm may have one advantage, namely that it seems to be easier to
parallelise.

\subsubsection*{Acknowledgements}

Ch.Z. is supported by CSE (Communications Security Establishment) and
MITACS (Mathematics of Information Technology and Complex Systems),
both from Canada.

\begin{appendix}
\section{Appendix: Detailed analysis of the success probability}
\label{FT A}

Here we analyse in some detail the success probability of the discrete
logarithm quantum algorithm when we use the usual quantum Fourier
transform of size $N=2^n$, as opposed to the ideal case which would
have prime size. The result is, that the algorithm has a probability
close to 1 of giving the right answer. Thus when looking at the
runtime we will assume that a single run is enough.

\subsection{Order finding algorithm (basis of factoring)}

We first consider the case of the order finding algorithm (section
\ref{order}) which is the basis of the factoring algorithm. The
discrete logarithm case is then simply a 2 dimensional version of
this.  Here we will use the eigenvalue estimation viewpoint introduced
by Kitaev \cite{kitaev} (see also \cite{cleve}). The advantage of this
viewpoint is, that the (mixed) state of the register which we
ultimately measure is explicitly written as a mixture of isolated
``peaks'' (thanks to Mike Mosca for pointing this out). In the usual
picture, which we used in section \ref{order}, we have the
diagonalised form of the mixed state (or, equivalently, we use the
Schmidt normal form between the entangled registers). But there we
have to worry about destructive interference between different peaks,
which makes the analysis a bit less nice.

So we want to find the order $r$ of a group element $\alpha$. Again we
do:
$$ \frac{1}{\sqrt{N}} \sum_x \ket{x} \quad \to \quad 
\frac{1}{\sqrt{N}} \sum_x \ket{x,\alpha^x} =
\frac{1}{\sqrt{N}} \sum_x \ket{x}~{U_\alpha^x} \ket{e} $$
Where $e$ is the neutral element and $U_\alpha$ is multiplication by
$\alpha$, thus $U_\alpha \ket{g} =\ket{\alpha g}$.  (Eigenvalue
estimation refers to the eigenvalues of $U_\alpha$.)  Now we write
$\ket{e}$ in terms of eigenstates of $U_\alpha$.  These $r$
eigenstates are easy to find:
$$\ket{\Psi_k} = \frac{1}{\sqrt{r}} \sum_{k'=0}^{r-1} {\omega_r}^{k
k'} \ket{\alpha^{k'}} \quad \mbox{with} \quad U_\alpha \ket{\Psi_k}=
{\omega_r}^{-k} \ket{\Psi_k} \quad \mbox{and} \quad \omega_r = e^{2
\pi i /r} $$
It is also easy to see that $\ket{e}$ is simply a uniform
superposition of these states:
$$ \ket{e} = \frac{1}{\sqrt{r}} \sum_k \ket{\Psi_k} $$
So the state of the quantum computer can be written as
$$ \frac{1}{\sqrt{N}} \sum_x \ket{x,\alpha^x} = \frac{1}{\sqrt{N}}
\sum_x \ket{x} ~\frac{1}{\sqrt{r}} \sum_k {\omega_r}^{-kx}
\ket{\Psi_k} $$
Now we apply the QFFT$_N$ to the first register to obtain:
$$ 
\frac{1}{\sqrt{r}} \sum_k \left( \sum_{x'} \frac{1}{N} \sum_x
{\omega_N}^{x x'} {\omega_r}^{-k x} \ket{x'} \right) \ket{\Psi_k}
$$
Because the $\ket{\Psi_k}$ are orthogonal, the state of the first
register alone can be viewed as a mixture of $r$ pure states, one for
each $k$. The probabilities associated with each of these pure states
are equal, namely $1/r$, as can be seen from the previous equation. By
summing the geometrical series in the sum over $x$ we get for these
(normalised) pure states:
\begin{eqnarray*}
\sum_{x'} \frac{1}{N} \sum_x {\omega_N}^{x x'} {\omega_r}^{-k x}
\ket{x'} 
 &=& \sum_{x'} \frac{1}{N} \frac{e^{2 \pi i N (x'/N -k/r)}-1}{e^{2 \pi i
 (x'/N -k/r)}-1} \ket{x'} =\\
= \sum_{x'} e^{i \phi(x')}
\frac{\sin(\pi (x' -k N/r))}{N \sin(\pi (x'/N -k/r))} \ket{x'} &=&
\sum_{x'} e^{i \phi(x')}
\frac{\sin(\pi (x' -x'_0))}{N \sin(\pi (x'-x'_0)/N)} \ket{x'}
\end{eqnarray*}
Where $\phi(x')$ is some (irrelevant) phase. We see that each of these
states is dominated by basis states $\ket{x'}$ with
$$x' \quad \approx \quad k \cdot N/r = x'_0 $$

\begin{figure}[tbp] 
\begin{center}
\includegraphics[scale = 1]{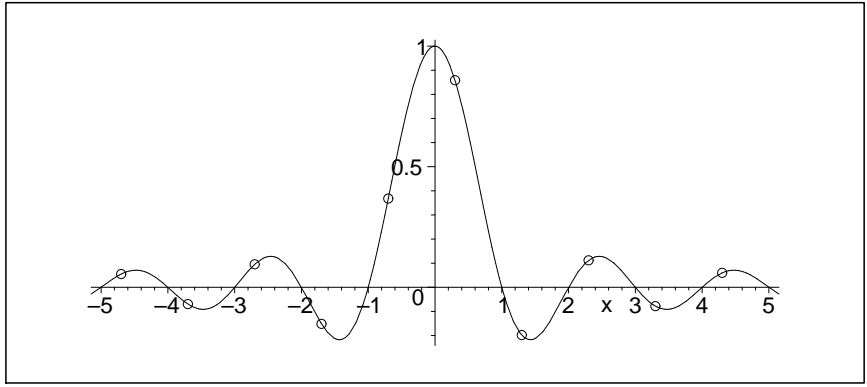}
\caption{\label{sinc} The function $\frac{\sin(\pi x)}{\pi x}$. Up to
an (irrelevant) phase, the amplitudes near a ``peak'' are given by
sampling this function at integer intervals, as indicated by the
circles.} 
\end{center}
\end{figure}

Thus each of the pure states corresponds to one ``smeared out'' peak
centered at $x'_0$.  Note that the argument of the sine in the
denominator is small. So the shape of the peak is approximately given
by the function $\sin(\pi x)/(\pi x)$ sampled at values for $x$ which
are integers plus some constant fractional offset, as plotted in
figure \ref{sinc}.

We are interested in the probability of observing a basis state no
farther away from the center $x'_0$ of the peak than, say $\Delta
x'$. How spread out the peak is, depends on the fractional offset. If
there is no offset, then we simply observe the central value with
probability 1. The largest spread occurs for offset $1/2$. (Then the
probabilities of the two closest basis states are each $4/\pi^2$.) The
chance of obtaining a state at distance $\Delta x'$ decreases as
$1/(\Delta x')^2$. So the probability of being away more than $\Delta
x'$ on either side is at most about $2/{\Delta x}$. Because the total
probability is normalised to 1, this tells us what the chance is of
coming within $\Delta x'$ of the central value.

\subsection{Discrete logarithm case}

The discrete logarithm case is analogous, actually it can be viewed as
a two dimensional version of the order finding algorithm. We have
$$ \sum_{x,y = 0}^{N-1} \ket{x,y,\alpha^x \beta^y} = 
\sum_{x,y} \ket{x,y,\alpha^{x+d y}} = 
\sum_{x,y} \ket{x,y} {U_\alpha}^{x+d y}
\frac{1}{\sqrt{q}} \sum_{k=0}^{q-1} \ket{\Psi_k}
$$
By applying a Fourier transform of size $N$ to each of the first two
registers we get
$$ 
\frac{1}{\sqrt{q}} \sum_k \left( \sum_{x', y'} 
~\frac{1}{N} \sum_x {\omega_N}^{x x'} {\omega_q}^{-k x} ~
~\frac{1}{N} \sum_y {\omega_N}^{y y'} {\omega_q}^{-d k y} ~ 
\ket{x',y'} \right) \ket{\Psi_k}
$$
Again we get a peak for each $k$, and each with the same
probability. The $x'$ and $y'$ values are independently distributed,
each as in the above 1-dimensional case. For $x'$ the ``central
value'' is $N k/q$ and for $y'$ it is $N d k/q$. To obtain the values
$k$ and $d k$ which we want, we multiply the observed $x',y'$ with
$q/N$ and round. Thus, if we chose $N$ ($=2^n$) sufficiently larger
than $q$, we are virtually guaranteed to obtain the correct values,
even if $x'$ and $y'$ are a bit off. Alternatively, we can try out
various integer values in the vicinity of our candidate $k$ and $d k$,
thus investing more classical post-processing to make the success
probability approach 1.

\section{Appendix: Bounding the number of cycles}
\label{AppB}
It was shown in section \ref{secNumSteps} that the number of cycles
 required to complete the Euclidean algorithm on inputs $p$ and $x$ is
\[
t(p,x) = 4\sum_{i=1}^{r} \lfloor \log_{2}(q_i) \rfloor +r
\]
where $q_1, q_2, \dots, q_r$ are the quotients in the Euclidean algorithm.
\begin{lemma}
If $p$ and $x$ are coprime integers such that $p >x \geq 1$ and $p >2$ then 
$t(p,x) \leq 4.5 \log_{2}(p)$.
\end{lemma}
\noindent \textbf{Proof:}
Assume by way of contradiction that there exist integers $(p,x)$ for which 
the lemma does not hold. 
Let $(p,x)$ be an input for which the number of Euclidean
iterations, $r$, 
is minimal subject to the condition that the lemma does not hold
(i.e. $t(p,x) > 4.5 \log_{2}(p)$, $p>2$, $p >x \geq 1$ 
 and $\gcd(p,x)=1$). 
Let $q_1,\dots, q_r$ be the quotients when the Euclidean algorithm is run on 
$(p,x)$.

We will now obtain a contradiction as follows. First, we show that 
if $t(p,x) > 4.5 \log_{2}(p)$ then the Euclidean algorithm with input $(p,x)$ 
will require at least three iterations (i.e. $r\geq 3$). Next, we show that 
if $t(p,x) > 4.5 \log_{2}(p)$ and the Euclidean algorithm run for 
two iterations on input $(p,x)$ returns the pair $(y,z)$ then $(y,z)$ also 
contradict the lemma. Since $(y,z)$ would contradict the lemma with 
fewer iterations than $(p,x)$ this contradicts the existence on $(p,x)$.
It is easily verified that the lemma holds provided $2< p \leq 15$ (simply 
calculate $t(p,x)$ for each of the possibilities). We can thus assume that 
$p \geq 16$. 

Recall that the Euclidean algorithm takes as input two integers $(a,b)$ and 
terminates when one of $a$ and $b$ is set to zero, at which point the other
integer will be $\gcd(a,b)$. An iteration of the Euclidean algorithm on 
$(a,b)$, with $a \geq b$, returns $(a-qb,b)$, where 
$q=\lfloor a/b \rfloor$. Note that since $\gcd(p,x)=1$ on this input 
 the algorithm will terminate with either $(1,0)$ or $(0,1)$.

Let us first prove that the Euclidean algorithm with input $(p,x)$ 
will require at least three iterations. 
Since neither $p$ nor $x$ is zero we know that $r \geq 1$. 
Suppose that $r=1$. Then the single iteration of the algorithm transforms 
$(p,x)$ to
$(p-q_1x,x) = (0,1)$. This implies that $x=1$ and $q_1=p$. Thus 
\[
t(p,x) = 4 \lfloor \log_{2}(p) \rfloor  +1 \leq 4.5 \log_{2}(p) 
 \hspace*{.5in} (\textrm{since } p >2 ) 
\]
which implies that $r\geq2$. Suppose that $r=2$. Then the two iterations of 
the algorithm would transform 
\[
(p,x) \rightarrow (p-q_1x,x) \rightarrow (p-q_1x,x-q_2 (p-q_1x)) = (1,0)
\]
This implies that $p-q_1x =1$ and $q_2 =x$. Thus $p -q_1q_2 =1$, which implies
that $\log_2(p) > \log_2(q_1) +  \log_2(q_2)$. Therefore  
\begin{eqnarray*}
t(p,x) &=& 4 \lfloor \log_{2}(q_1) \rfloor + 4 \lfloor \log_{2}(q_2) \rfloor + 2 \\
 & \leq & 4 \lfloor \log_{2}(q_1) + \log_{2}(q_2) \rfloor +2 \\
 & < & 4 \lfloor \log_{2}(p) \rfloor +2 \\
&\leq & 4.5 \log_{2}(p) \hspace*{.5in} (\textrm{since } p  \geq 16 ) 
\end{eqnarray*}
and we have that $r\geq 3$. Note that we now know $x\neq 1,2$, 
$p-q_1x \neq 1$ and $x-q_2(p-q_1x) \neq 0$ since any of these would imply
$r \leq 2$.

We shall now establish that $q_1 \in \{1,2\}$.
After the first iteration of the Euclidean algorithm the problem is 
reduced to running the algorithm on $(p-q_1x,x)$, for which
 the quotients will be  $q_2,\dots, q_r$. Since $xq_1 \leq p$ 
we have that $\log_{2}(p) \geq \log_{2}(x) + \log_{2}(q_1)$. Therefore
\begin{eqnarray*}
t(x,p-q_1x) &=& 4\sum_{i=2}^{r} \lfloor \log_{2}(q_i) \rfloor +r -1 \\
&=& t(p,x)  - (4 \lfloor \log_{2}(q_1) \rfloor +1) \\
&>& 4.5 \log_{2}(p) - (4 \lfloor \log_{2}(q_1) \rfloor +1) \\
&\geq&  4.5 \log_{2}(x) + 4.5\log_{2}(q_1) -4 \lfloor \log_{2}(q_1) \rfloor-1\\
&\geq&  4.5 \log_{2}(x) \hspace*{.5in} (\textrm{if } q_1 \geq 3) 
\end{eqnarray*}
Thus if $q_1 \geq 3$ then $t(x,p-q_1x) > 4.5\log_{2}(x)$, $x >2$ and 
$x > p-q_1x \geq 1$, but this would contradict the minimality of $r$.
Therefore $q_1 \in \{1,2\}$.

After two iterations of the Euclidean algorithm on $(p,x)$ the problem has 
been reduced to running the algorithm on $(p-q_1x, x-q_2(p-q_1x))$. We will 
now show that the lemma does not hold for $(p-q_1x, x-q_2(p-q_1x))$.
This will contradict the minimality of $r$ and thus the 
existence of $(p,x)$. 
To do this, we must first show that $p-q_1x >2$ and that 
$p-q_1x > x-q_2(p-q_1x) \geq 1$ (so that the lemma applies). As discussed 
above, since $r \geq 3$ we know that $p-q_1x >1$ and that
 $p-q_1x > x-q_2(p-q_1x) \geq 1$, thus we need only show that $p-q_1x \neq 2$.

Suppose that $p-q_1x=2$.  Since $q_1 \in \{1,2\}$ 
 either $p=x+2$ or 
$p=2x+2$. Since $\gcd(p,x)=1$ this implies that $x$ is odd and that 
the Euclidean algorithm will proceed as follows
\[
(p,x) \rightarrow (2,x) \rightarrow (2,1) \rightarrow (0,1)
\]
Thus $r=3$, $q_2= (x-1)/2$, $q_3=2$ and 
\begin{eqnarray*}
t(p,x) &=& 4 \lfloor \log_{2}(q_1) \rfloor + 4  \lfloor \log_{2}((x-1)/2) \rfloor +  4 \lfloor \log_{2}(2) \rfloor +3 \\
&=& 4 \lfloor \log_{2}(q_1x-q_1)) \rfloor + 3 \\
& \leq & 4.5 \log_2(p=q_1x+2)
\end{eqnarray*}
where the last line follows by checking the values for $q_1 \in \{1,2\}$ and 
$x <64$ and noting that $4.5\log_2(q_1x) > 4\log_2(q_1x) +3$ when $x\geq 64$.
This would contradict the fact that the lemma doesn't hold for $(p,x)$, thus 
$p-q_1x \neq 2$. 

Now to complete the proof we need only show that 
$t(p-q_1x, x-q_2(p-q_1x)) > 4.5 \log_2(p-q_1x)$.
Let $x=cp$, so $p-q_1x=(1-q_1c)p$ with $1> 1-q_1c > 0$.
By the Euclidean algorithm we know that
$x \geq q_2(p-q_1x)$ and thus    
\[
\log_2(x) = \log_{2}(p) + \log_{2}(c) \geq \log_{2}(p) + \log_{2}(1-q_1c) + \log_{2}(q_2)
\]
Therefore  $\log_{2}(c/(1-q_1c)) \geq \log_{2}(q_2)$, which implies $1-q_1c \leq 1/(1+q_1q_2)$. 
This in turn implies that  $\log_{2}(p-q_1x) = \log_{2}(p) + \log_{2}(1-q_1c) \leq \log_{2}(p) - \log_{2}(1 + q_1q_2)$.
Hence 
\begin{eqnarray*}
t(p-q_1x, x-q_2(p-q_1x)) &=& 4\sum_{i=3}^{r} \lfloor \log_{2}(q_i) \rfloor +r -2 \\
&=& t(p,x)  - (4 \lfloor \log_{2}(q_2) \rfloor +4 \lfloor \log_{2}(q_1) \rfloor +2) \\
&>& 4.5 \log_{2}(p) -(4\lfloor \log_{2}(q_2) \rfloor +4\lfloor \log_{2}(q_1) \rfloor +2)\\
&\geq&  4.5 \log_{2}(p-q_1x) + Z(q_1,q_2)
\end{eqnarray*}
where $Z(q_1,q_2)=4.5\log_{2}(1+q_1q_2) - (4\lfloor \log_{2}(q_2) \rfloor +4\lfloor \log_{2}(q_1) \rfloor +2)$. 

If $q_1=1$ then $Z(q_1,q_2) =4 .5\log_{2}(1+q_2) - (4\lfloor \log_{2}(q_2) \rfloor+2)$.
It is easy to check that $Z(1,q_2)$ is non-negative when 
$q_2 \in \{1,\dots,14\}$ and if $q_2 \geq 15$ then $Z(1,q_2) > .5\log_{2}(1+q_2) -2 \geq 0$. Therefore $Z(q_1,q_2) \geq 0$ when $q_1=1$.

If $q_1=2$ then $Z(q_1,q_2) =4.5\log_{2}(1+2q_2)- (4\lfloor \log_{2}(q_2) \rfloor+6)$.
It is easy to check that $Z(2,q_2)$ is non-negative when $q_2 \in \{1,\dots, 7\}$ and if $q_2 \geq 8$ then 
\begin{eqnarray*}
Z(2,q_2) &=& 4.5\log_{2}(1+2q_2)- (4\lfloor \log_{2}(q_2) \rfloor+6) \\
 &>& 4.5(\log_{2}(q_2) +1) - (4\lfloor \log_{2}(q_2) \rfloor+6) \\
&\geq& .5\log_{2}(q_2) -1.5 \\
&\geq& 0
\end{eqnarray*}
 Therefore $Z(q_1,q_2) \geq 0$ when $q_1=2$.

Thus $Z(q_1,q_2) \geq 0$ and we have that 
$t(p-q_1x, x-q_2(p-q_1x)) > 4.5 \log_{2}(p-q_1x)$. This 
contradict the minimality of $r$ and thus the existence of $(p,x)$. 
Therefore the lemma holds.\hfill $\Box$\\

Note that $t(4,1) = 9 = 4.5\log_{2}(p)$ and thus the bound is
tight. It is also worth noting that $t(2,1) = 5 = 5 \log_{2}(p)$ which
is why the requirement $p>2$ was included in the lemma.

\end{appendix}

\end{document}